\documentclass[prd,twocolumn,showpacs,amsfonts,amssymb]{revtex4}

\usepackage{graphicx}

\usepackage{bm}

\newcommand{\beq}{\begin{equation}}
\newcommand{\eeq}{\end{equation}}
\newcommand{\beqs}{\begin{eqnarray}}
\newcommand{\eeqs}{\end{eqnarray}}
\newcommand{\lsim}{\mathrel{\raisebox{-
.6ex}{$\stackrel{\textstyle<}{\sim}$}}}
\newcommand{\gsim}{\mathrel{\raisebox{-
.6ex}{$\stackrel{\textstyle>}{\sim}$}}}

\begin{document}

\title{Scheme Transformations in the Vicinity of an Infrared Fixed Point} 

\author{Thomas A. Ryttov$^a$ and Robert Shrock$^b$}

\affiliation{(a) \ Jefferson Physical Laboratory \\ 
Physics Department, Harvard University,  \\
Cambridge, MA  02138}

\affiliation{(b) \ C. N. Yang Institute for Theoretical Physics \\
Stony Brook University \\
Stony Brook, NY 11794 }

\begin{abstract}

We analyze the effect of scheme transformations in the vicinity of an exact or
approximate infrared fixed point in an asymptotically free gauge theory with
fermions.  We show that there is far less freedom in carrying out such scheme
transformations in this case than at an ultraviolet fixed point.  We construct
a transformation from the $\overline{MS}$ scheme to a scheme with a vanishing
three-loop term in the $\beta$ function and use this to assess the scheme
dependence of an infrared fixed point in SU($N$) theories with fermions.
Implications for the anomalous dimension of the fermion bilinear operator are
also discussed.

\end{abstract}

\pacs{11.15.-q,11.10.Hi,12.60.-i}

\maketitle

The evolution of an asymptotically free gauge theory from the weakly coupled
ultraviolet (UV) regime to the infrared (IR) regime is of fundamental
interest. Here we study this evolution for a theory with gauge group $G$ and a
given content of massless fermions.  We focus mainly on vectorial gauge
theories (VGT) \cite{fm}, but also remark on chiral gauge theories
($\chi$GT). The UV to IR evolution is determined by the renormalization group
$\beta$ function of the theory, which describes the dependence of $g \equiv
g(\mu)$, the running gauge coupling, on the Euclidean momentum scale, $\mu$. We
define $\alpha=g^2/(4\pi)$, $a = \alpha/(4\pi)$, and $\beta_\alpha \equiv
d\alpha/dt$, where $t=\ln \mu$.  This has the series expansion
\beq
\beta_\alpha = -2\alpha \sum_{\ell=1}^\infty b_\ell \, a^\ell = 
 -2\alpha \sum_{\ell=1}^\infty \bar b_\ell \, \alpha^\ell \ , 
\label{beta}
\eeq
where $\bar b_\ell = b_\ell/(4\pi)^\ell$.  The coefficients $b_1$ and $b_2$
were calculated in \cite{b1} and \cite{b2}, respectively.  The asymptotic
freedom (AF) property is the condition $b_1 > 0$, which we assume.  As
discussed further below, the $b_\ell$ for $\ell=1,2$ are independent of the
scheme used for regularization and renormalization, while $b_\ell$ with $\ell
\ge 3$ are scheme-dependent \cite{gross75}.  One scheme involves dimensional
regularization \cite{dimreg} and minimal subtraction ($MS$) of the poles at
dimension $d=4$ in the resultant Euler $\Gamma$ functions \cite{ms}.  The
modified minimal subtraction ($\overline{MS}$) scheme also subtracts certain
related constants \cite{msbar}. Calculations of $b_3$ and $b_4$ in the
$\overline{MS}$ scheme were given in \cite{b3,b4}. Just as the calculation of
$b_1$ and demonstration that $b_1 > 0$ was pivotal for the development of QCD,
the computation of $b_\ell$ for $\ell=2,3,4$ has been important in fits to
$\alpha_s(Q)$ \cite{bethke}.  In the vicinity of the UV fixed point (UVFP) at
$\alpha=0$, one can carry out a scheme transformation that renders three- and
higher-loop terms zero \cite{thooft77}.  Considerable work has been done on
scheme (and related scale) transformations that reduce higher-order corrections
in QCD \cite{brodskyst}.

Naively, one might think that there is a similarly great freedom in performing
scheme transformations at an (exact or approximate) IRFP.  Here we show that,
on the contrary, there is much less freedom in constructing acceptable scheme
transformations at an IRFP than at a UVFP, and we analyze constraints at an
IRFP. We construct an example of a scheme transformation that satisfies these
constraints, and we apply it to assess scheme-dependence of the value of an
IRFP.

We first recall some background.  In a non-Abelian gauge theory with no
fermions or only a few fermions, $b_2$ has the same positive sign as $b_1$, so
$\beta$ has no (perturbative) IR zero for $\alpha \ne 0$ \cite{nonpertzero}.
With a sufficient increase in the content of fermions, $b_2$ reverses sign,
while $b_1$ is still positive, so the two-loop $\beta$ function has a zero at
\beq
\alpha_{IR,2\ell} = - \frac{4\pi b_1}{b_2} \ , 
\label{alfir_2loop}
\eeq
which is physical for $b_2 < 0$.  This zero plays an important role in the UV
to IR evolution of the theory \cite{b2,bz}. If $\alpha_{IR,2\ell}$ is large
enough, then, as $\mu$ decreases through a scale denoted $\Lambda$, the gauge
interaction grows strong enough to produce a bilinear fermion condensate in the
most attractive channel (MAC) with attendant spontaneous chiral symmetry
breaking (S$\chi$SB) and dynamical generation of effective masses for the
fermions involved \cite{conf}. In a one-gluon exchange approximation to the
Dyson-Schwinger equation for the fermion propagator in a VGT, this occurs as
$\alpha$ increases through a value $\alpha_{cr}$ given by $\alpha_{cr} C_f \sim
O(1)$ \cite{wtc,chipt,dscor}. In a chiral gauge theory this breaks the gauge
symmetry, while in the vectorial case, the MAC is $R \times \bar R \to 1$,
preserving the gauge symmetry \cite{tc}. Since the fermions that have gained
dynamical masses are integrated out in the low-energy effective field theory
below $\Lambda$, the $\beta$ function changes, and the theory flows away from
the original IRFP, which is thus only approximate.  However, if
$\alpha_{IR,2\ell}$ is sufficiently small, as is the case with a large enough
(AF-preserving) fermion content, then the theory evolves from the UV to the IR
without any S$\chi$SB. In this case the theory has an exact IRFP.  For a given
$G$ and $N_f$ (massless) fermions in a representation $R$, the critical value
of $N_f$ beyond which the theory flows to the IR conformal phase is denoted
$N_{f,cr}$.  As $N_f$ increases, $\alpha_{IR,2\ell}$ decreases, and $N_{f,cr}$
is the value at which $\alpha_{IR,2\ell}$ decreases through $\alpha_{cr}$.
Lattice simulations have been used to estimate $N_{f,cr}$ \cite{lgt}.

Since $\alpha_{IR,2\ell}$ is $\sim O(1)$, especially in the quasi-conformal
case where $N_f \lsim N_{f,cr}$, there are significant corrections to the
two-loop results from higher-loop terms in $\beta$.  These motivate one to
calculate these corrections to three- and four-loop order, and we have done
this in the $\overline{MS}$ scheme \cite{bvh,bfs} (see also \cite{ps}, which
agrees with \cite{bvh}).  Because of the scheme-dependence of $b_n$ for $n \ge
3$, the value of $\alpha_{IR,n\ell}$ calculated to finite order $n \ge 3$ is
scheme-dependent.  It is important to assess this scheme dependence and the
resultant uncertainties in the value of the (exact or approximate) IRFP.  We
address this task here.  Besides its intrinsic field-theoretic interest, this
is important for ongoing studies of quasi-conformal theories.  These have a
gauge coupling that gets large but runs slowly over a long interval of $\mu$
\cite{wtc,chipt}, as occurs naturally due to an approximate IRFP \cite{chipt}.
Moreover, the UV to IR flow of a $\chi$GT and the associated sequential gauge
symmetry breaking are important in certain approaches to physics beyond the
Standard Model \cite{as}.

A scheme transformation (ST) is a map between $\alpha$ and $\alpha'$. It will
be convenient to write this as 
\beq
a = a' f(a') \ . 
\label{aap}
\eeq
To keep the UV properties the same, one requires $f(0) = 1$.  We consider STs
that are analytic about $a=a'=0$ \cite{nonan} and hence can be expanded in the
form
\beq
f(a') = 1 + \sum_{s=1}^{s_{max}} k_s (a')^s =
        1 + \sum_{s=1}^{s_{max}} \bar k_s (\alpha')^s \ ,
\label{faprime}
\eeq
where the $k_s$ are constants, $\bar k_s = k_s/(4\pi)^s$, and $s_{max}$ may be
finite or infinite. Hence, the Jacobian $J=da/da'$ satisfies $J=1$ at
$a=a'=0$. We have
\beq
\beta_{\alpha'} \equiv \frac{d\alpha'}{dt} = \frac{d\alpha'}{d\alpha} \, 
\frac{d\alpha}{dt} = J^{-1} \, \beta_{\alpha} \ . 
\label{betaap}
\eeq
This has the expansion 
\beq
\beta_{\alpha'} = -2\alpha' \sum_{\ell=1}^\infty b_\ell' (a')^\ell =
-2\alpha' \sum_{\ell=1}^\infty \bar b_\ell' (\alpha')^\ell \ ,
\label{betaprime}
\eeq
where $\bar b'_\ell = b'_\ell/(4\pi)^\ell$.  Given the equality of
Eqs. (\ref{betaap}) and (\ref{betaprime}), one can solve for the $b_\ell'$ in
terms of the $b_\ell$ and $k_s$. This leads to the well-known result that
$b_\ell' = b_\ell$ for $\ell=1,2$ \cite{gross75}, i.e., that the one- and
two-loop terms in $\beta$ are scheme-independent \cite{aapap}. We note that the
scheme-invariance of $b_2$ assumes that $f(a')$ is gauge-invariant. This
is evident from the fact that in the momentum subtraction (MOM) scheme, $b_2$
is actually gauge-dependent \cite{mom} and is not equal to $b_2$ in the
$\overline{MS}$ scheme.  We restrict our analysis here to gauge-invariant STs
and to schemes, such as $\overline{MS}$, where $b_2$ is gauge-invariant.

In order to assess scheme-dependence of an IRFP, we have calculated the
relations between the $b'_\ell$ and $b_\ell$ for higher $\ell$. For example,
for $\ell=3,4,5$ we obtain
\beq
b_3' = b_3 + k_1b_2+(k_1^2-k_2)b_1 \ , 
\label{b3prime}
\eeq
\beq
b_4' = b_4 + 2k_1b_3+k_1^2b_2+(-2k_1^3+4k_1k_2-2k_3)b_1 \ , 
\label{b4prime}
\eeq
\beqs
b_5' & = & b_5+3k_1b_4+(2k_1^2+k_2)b_3+(-k_1^3+3k_1k_2-k_3)b_2 \cr\cr
     & + & (4k_1^4-11k_1^2k_2+6k_1k_3+4k_2^2-3k_4)b_1 \ .
\label{b5prime}
\eeqs
In general, in the coefficients of the terms $b_n$ entering in the expression
for $b'_\ell$, the sum of the subscripts of the $k_s$s is equal to $\ell-n$
with $1 \le n \le \ell-1$, and the products of the various $k_s$s correspond to
certain partitions of $\ell-n$. A corollary is that the only $k_s$s that appear
in the formula for $b_\ell'$ are the $k_s$s with $1 \le s \le \ell-1$.
However, because of cancellations, in the expression for $b_\ell'$ for even
$\ell$, the coefficient of $b_n$ does not contain all of the terms
corresponding to the partitions of $\ell-n$.  For example, in $b'_2$, there is
no $k_1b_1$ term and in $b'_4$, the coefficient of $b_2$ does not contain
$k_2$.

In order to be physically acceptable, this transformation must satisfy several
conditions, $C_i$.  For finite $s_{max}$, Eq. (\ref{aap}) is an algebraic
equation of degree $s_{max}+1$ for $\alpha'$ in terms of $\alpha$.  We require
that at least one of the $s_{max}+1$ roots must satisfy these conditions. These
are as follows: $C_1$: the ST must map a real positive $\alpha$ to a real
positive $\alpha'$, since a map taking $\alpha > 0$ to $\alpha'=0$ would be
singular, and a map taking $\alpha > 0$ to a negative or complex $\alpha'$
would violate the unitarity of the theory.  $C_2$: the ST should not map a
moderate value of $\alpha$, for which perturbation theory may be reliable, to a
value of $\alpha'$ that is so large that perturbation theory is
unreliable. $C_3$: \ $J$ should not vanish in the region of $\alpha$ and
$\alpha'$ of interest, or else there would be a pole in Eq. (\ref{betaap}).
The existence of an IR zero of $\beta$ is a scheme-independent property of an
AF theory, depending (insofar as perturbation theory is reliable) only on the
condition that $b_2 < 0$.  Hence, $C_4$: an ST must satisfy the condition that
$\beta_\alpha$ has an IR zero if and only if $\beta_{\alpha'}$ has an IR zero.
These four conditions can always be satisfied by scheme transformations used to
study the UVFP at $\alpha=\alpha'=0$ and hence in applications to perturbative
QCD calculations, since the gauge coupling is small (e.g., $\alpha_s(m_Z) =
0.118$), and one can choose the $k_s$ to have small magnitudes.

However, we stress that these conditions are not automatically satisfied, and
are significant constraints, in the analysis of an (exact or approximate)
IRFP.  To show this, we first exhibit an apparently reasonable ST that
satisfies $C_1$ and $C_3$ but fails $C_2$ and $C_4$.  This is the map (with
$s_{max}=\infty$) \cite{tanhgennote} 
\beq
\alpha=\tanh(\alpha')
\label{tanh}
\eeq
with the inverse $\alpha' = (1/2)\ln[(1+\alpha)/(1-\alpha)]$ and Jacobian
$J=1/\cosh^2(\alpha')$. This ST is acceptable at a UVFP. But at an IRFP, it can
easily happen that $\alpha_{IR,2\ell} > 1$, in which case this ST yields a
complex, unphysical $\alpha'$. For example (see Table III in \cite{bvh}) for
$G={\rm SU}(2)$ with $N_f=8$ fermions in the fundamental representation, 
$\alpha_{IR,2\ell}=1.26$ and for SU(3) with $N_f=11$, $\alpha_{IR,2\ell}=1.23$.

To exhibit another type of pathology that can arise at an IRFP, but not a UVFP,
consider an ST with $s_{max}=2$ and, for simplicity, $k_1=0$, viz.,
\beq
a=a'[1+k_2(a')^2] 
\label{aapsmax2k1zero}
\eeq
with a moderate value of $|k_2|$.  This is a cubic equation for $a'$ in terms
of $a$, and, by continuity arguments, in the vicinity of the UVFP, it is
guaranteed that this cubic yields a root that satisfies $C_1$-$C_4$.  But the
situation is different at an IRFP. Consider sufficiently large $N_f$ that $b_2
< 0$, so there is a two-loop zero of $\beta$, at the value (\ref{alfir_2loop}).
For a given $G$ and $R$, as $N_f$ increases from 0, $b_2$ decreases through
positive values and vanishes, becoming negative, as $N_f$ increases through the
value $N_{f,b2z}=17C_A^2/[2T_f(5C_A+3C_f)]$ (which is always less than the
value $N_{f,b1z}=11C_A/(4T_f)$ at which $b_1$ turns negative and AF is lost)
\cite{casimir,nfreal}.  The two-loop IR zero of $\beta$ is thus present for
$N_f$ in the interval $I$ defined by $N_{f,b2z} < N_f < N_{f,b1z}$. Now with
$N_f \in I$, let us investigate the ST (\ref{aapsmax2k1zero}).  The condition
$b'_3=0$ is then a linear equation for $k_2$, with the solution $k_2 =
b_3/b_1$.  To guarantee that this ST satisfies $C_1$, we require $1+k_2(a')^2 >
0$, i.e., $1+(b_3/b_1)(a')^2 > 0$.  This must be satisfied, in particular, in
the vicinity of the two-loop IR zero of $\beta$, so substituting the
(scheme-independent) $a_{IR,2\ell}=a'_{IR,2\ell}=-b_1/b_2$ from Eq.
(\ref{alfir_2loop}), we obtain the inequality
\beq
1 + \frac{b_1b_3}{b_2^2} > 0 \ . 
\label{conditionc2ir}
\eeq
But this inequality is not, in general, satisfied.  This can be seen by
substituting explicit values of $b_\ell$ from Table I of \cite{bvh} for
$G={\rm SU}(N)$ and $N_f$ fermions in the fundamental representation, for
example.

We proceed to construct and study an ST that does satisfy our constraints and
provides a measure of the scheme dependence of the value of the IR zero of
$\beta$ that we calculated in \cite{bvh} up to four-loop order in the
$\overline{MS}$ scheme.  We assume $N_f \in I$, so a two-loop IR zero of
$\beta$ exists. Starting in this $\overline{MS}$ scheme, we construct an ST
with $s_{max}=1$ that yields $b'_3=0$.  Eq. (\ref{aap}) reads $a=a'(1+k_1
a')$. Solving this for $a'$, or equivalently, $\alpha'$, we have, formally, two
solutions,
\beq
\alpha'_{\pm} = \frac{1}{2\bar k_1} (-1 \pm \sqrt{1+4 \bar k_1 \alpha} \ ) \ .
\label{alfprimesmax1}
\eeq
Only $\alpha'_+$ is acceptable, since only this solution has $\alpha \to
\alpha'$ as $\alpha \to 0$. For $\alpha'$ to be real, it is necessary that
$\bar k_1 > -1/(4\alpha)$. Solving the equation
$b'_3=0$ for $k_1$, we get, formally, two solutions,
\beq
k_{1p}, \ k_{1m} = \frac{1}{2b_1} \Big [ -b_2 \pm \sqrt{b_2^2-4b_1b_3} \ 
\Big ] \ , 
\label{b3pzerosolclass1}
\eeq
where $(p,m)$ refer to $\pm$.  We will focus on $G={\rm SU}(N)$ with fermions
in the fundamental and adjoint representation. The discriminant $b_2^2-4b_1b_3
> 0$ satisfies the requirement of being non-negative here. The solution
$k_{1m}$ must be discarded because it leads to $\alpha$ and $\alpha'$ having
opposite signs for some $N_f \in I$. We thus choose the solution $k_{1p}$.  We
denote this as the $S_1$ scheme, i.e.,
\beq
S_1: \quad\quad a=a'(1+k_{1p}a') \ .
\label{sprime}
\eeq
By construction, since $b'_3=0$ in this scheme, the three-loop zero of
$\beta_{\alpha'}$ is equal to the two-loop zero,
$\alpha'_{IR,3\ell}=\alpha'_{IR,2\ell} = \alpha_{IR,2\ell} = -4\pi b_1/b_2$
\cite{aapap}.  At the four-loop level, the IR zero is given by the physical
(smallest positive) solution of the cubic $b_1 + b_2 a' + b_4' (a')^3 = 0$,
with $b'_4$ given by Eq. (\ref{b4prime}) with $k_1=k_{1p}$ and $k_2=k_3=0$.

We have calculated the resultant $\alpha'_{IR,n\ell}$ in the $S_1$ scheme up to
$(n=4)$-loop level. In Table \ref{betazero_fund} we list values of the $n$-loop
IR zero, $\alpha'_{IR,n\ell}$ for $n=2,3,4$ for relevant $N_f$, with fermions
in the fundamental representation and several values of
$N$.  For comparison we also include the values of
$\alpha_{IR,n\ell}$ for $n=3,4$ in the $\overline{MS}$ scheme from \cite{bvh}.
We have carried out the analogous calculations for fermions in the adjoint
representation of SU($N$). Here, $N_{f,b1z}=11/4$ and $N_{f,b2z}=17/16$, so the
only physical, integer value of $N_f \in I$ is $N_f=2$.  SU(2) models with
$N_f=2$ adjoint fermions have been of recent interest \cite{sanrev}. We list
our results in Table \ref{betazero_adj}. For both of these cases we find 
that $\alpha'_{IR,3\ell} > \alpha_{IR,3\ell,\overline{MS}}$ and
$\alpha'_{IR,4\ell} < \alpha_{IR,4\ell,\overline{MS}}$. 

The anomalous dimension $\gamma_m$ describes the scaling of a fermion bilinear
and the running of a dynamically generated fermion mass in the phase with
S$\chi$SB. It plays an important role in technicolor theories, via the
renormalization group factor $\eta = \exp[\int dt \, \gamma_m(\alpha(t))]$ that
can enhance dynamically generated Standard-Model fermion masses.  In the
(conformal) non-Abelian Coulomb phase, the IR zero of $\beta$ is exact,
although a calculation of it to a finite-order in perturbation theory is only
approximate, and $\gamma_m$ evaluated at this IRFP is exact. In the phase with
S$\chi$SB, where an IRFP, if it exists, is only approximate, $\gamma_m$ is an
effective quantity describing the running of a dynamically generated fermion
mass for the evolution of the theory near this approximate IRFP.  In \cite{bvh}
we evaluated $\gamma_m$ to three- and four-loop order at the IR zero of $\beta$
calculated to the same order and showed that these higher-loop results were
somewhat smaller than the two-loop evaluation.  In both the conformal and
nonconformal phases it is important to assess the scheme-dependence of
$\gamma_m$ when calculated to finite order.  $\gamma_m$ is defined as
$\gamma_m = d\ln Z_m/dt$, where $Z_m$ is the corresponding renormalization
constant.  This has the expansion $\gamma_m = \sum_{\ell=1}^\infty \bar c_\ell
\, \alpha^\ell$ with $\bar c_\ell$ calculated up to $\ell=4$ order in the
$\overline{MS}$ scheme \cite{gamma4}.  Under the general ST (\ref{aap}), $c_1$
is invariant, while the $c_\ell$ with $\ell \ge 2$ change.  With $Z_m(\alpha) =
Z_m'(\alpha')F_m(\alpha')$,
\beq
\gamma_m(\alpha) = \gamma_m'(\alpha') + \frac{d\alpha'}{dt} \, 
\frac{d\ln F_m}{d\alpha'} = \gamma_m'(\alpha') + \beta_{\alpha'} \, 
\frac{d\ln F_m}{d\alpha'} \ . 
\label{gamrel}
\eeq
Hence, at a zero of $\beta_{\alpha'}$, $\gamma_m(\alpha) = \gamma_m'(\alpha')$
\cite{gross75}. Although $\gamma_m$ calculated to all orders is invariant under
an ST at a zero of $\beta$, in particular an exact IRFP, our present results
with the $\overline{MS}$ and $S_1$ schemes show that $\alpha_{IR,n\ell}$ and
$\gamma_{m,n\ell}(\alpha_{IR,n\ell})$ still exhibit significant
scheme-dependence up to ($n=4$)-loop order. This is understandable, since the
relevant IRFP occurs at $\alpha \sim O(1)$.

It is also of interest to consider STs that are not
designed to render any $b'_\ell = 0$. Accordingly, we have also done
calculations with one-parameter STs having $s_{max}=\infty$ and exactly known
inverses, such as
\beq
a=\frac{\tanh(ra')}{r}
\label{tanhgen}
\eeq
and $a=(1/r)\sinh(ra')$, where $r$ is a positive constant.  For these we can
vary the effect of the transformation by varying $r$ from $r << 1$ to values $r
\gsim 1$.  These STs provide a further measure of the scheme-dependence of an
IRFP \cite{scc2}. 

This research was partially supported by a Sapere Aude Grant (TAR) and 
NSF grant NSF-PHY-09-69739 (RS).

\newpage

\begin{table}
\caption{\footnotesize{Values of the IR zeros of $\beta_{\alpha}$ in the
$\overline{MS}$ scheme and $\beta_{\alpha'}$ in the $S_1$ scheme, for an
SU($N$) theory with $N_f$ fermions in the fundamental representation, for
$N=2,3,4$, calculated to $n$-loop order and denoted as
$\alpha_{IR,n\ell,\overline{MS}}$ and $\alpha'_{IR,n\ell}$.  Here,
$\alpha_{IR,2\ell,\overline{MS}}=\alpha'_{IR,2\ell}$ is scheme-independent, so
we denote it simply as $\alpha_{IR,2\ell}$. In the $S_1$ scheme,
$\alpha'_{IR,3\ell}=\alpha'_{IR,2\ell}=\alpha_{IR,2\ell}$.}}
\begin{center}
\begin{tabular}{|c|c|c|c|c|c|} \hline\hline
$N$ & $N_f$ & $\alpha_{IR,2\ell}$ & $\alpha_{IR,3\ell,\overline{MS}}$ 
&                                     $\alpha_{IR,4\ell,\overline{MS}}$ & 
                $\alpha'_{IR,4\ell}$ \\
\hline
 2  &  7  &  2.83   & 1.05   & 1.21   & 0.640  \\
 2  &  8  &  1.26   & 0.688  & 0.760  & 0.405  \\
 2  &  9  &  0.595  & 0.418  & 0.444  & 0.2385  \\
 2  & 10  &  0.231  & 0.196  & 0.200  & 0.109  \\
 \hline
 3  & 10  &  2.21   & 0.764  & 0.815  & 0.463  \\
 3  & 11  &  1.23   & 0.578  & 0.626  & 0.344  \\
 3  & 12  &  0.754  & 0.435  & 0.470  & 0.254  \\
 3  & 13  &  0.468  & 0.317  & 0.337  & 0.181  \\
 3  & 14  &  0.278  & 0.215  & 0.224  & 0.121  \\
 3  & 15  &  0.143  & 0.123  & 0.126  & 0.068  \\
 3  & 16  &  0.0416 & 0.0397 & 0.0398 & 0.0215 \\
\hline
 4  & 13  &  1.85   & 0.604  & 0.628  & 0.365  \\
 4  & 14  &  1.16   & 0.489  & 0.521  & 0.293  \\
 4  & 15  &  0.783  & 0.397  & 0.428  & 0.235  \\
 4  & 16  &  0.546  & 0.320  & 0.345  & 0.187  \\
 4  & 17  &  0.384  & 0.254  & 0.271  & 0.146  \\
 4  & 18  &  0.266  & 0.194  & 0.205  & 0.110  \\
 4  & 19  &  0.175  & 0.140  & 0.145  & 0.0785 \\
 4  & 20  &  0.105  & 0.091  & 0.092  & 0.050  \\
 4  & 21  &  0.0472 & 0.044  & 0.044  & 0.023  \\
\hline\hline
\end{tabular}
\end{center}
\label{betazero_fund}
\end{table}

\begin{table}
\caption{\footnotesize{Values as in Table \ref{betazero_fund}, but for 
$N_f=2$ fermions in the adjoint representation of SU($N$).}}
\begin{center}
\begin{tabular}{|c|c|c|c|c|} \hline\hline
$N$ & $\alpha_{IR,2\ell,adj}$ & $\alpha_{IR,3\ell,adj,\overline{MS}}$ &
$\alpha_{IR,4\ell,adj,\overline{MS}}$ & $\alpha'_{IR,4\ell,adj}$ \\ \hline
 2  &  0.628  & 0.459   & 0.450  & 0.258 \\
 3  &  0.419  & 0.306   & 0.308  & 0.173 \\
 4  &  0.314  & 0.2295  & 0.234  & 0.130 \\
\hline\hline
\end{tabular}
\end{center}
\label{betazero_adj}
\end{table}


\begin{thebibliography}{99}
\bibitem{fm}
%
It is straightforward to generalize our analysis to finite fermion masses 
in a vectorial gauge theory.
\bibitem{nonan}
%
One can generalize this to certain STs with $f(a')$ functions that are 
finite but nonanalytic at $a'=0$, such as $f(a') = [1+\sum_{s=1}^{s_{max}} k_s
(a')^s ][1 + \kappa e^{-\nu/a'}]$, where $\kappa$ and $\nu$ are (real)
constants and $\nu > 0$.  Eqs. (\ref{b3prime})-(\ref{b5prime}) and the
analogous relations for higher-$s$ continue to hold
for these scheme transformations.
\bibitem{aapap}
%
Hence, if there is an IR zero in the two-loop $\beta_\alpha$, at 
$\alpha_{IR,2\ell}$ given by (\ref{alfir_2loop}), then there is also an 
IR zero in the two-loop $\beta_{\alpha'}$ at the same value of $\alpha'$,
This is consistent with the fact that (\ref{aap}) in general maps 
$a'=-b_1/b_2$ to $a \ne -b_1/b_2$, since (\ref{aap}) is an exact
result, whereas the equality of two-loop IR zero values holds for the
truncations of $\beta$ and $\beta'$ to two-loop order.
\bibitem{thooftscheme}
%
In particular, near a UVFP, if we set $s_{max}=\infty$, we can choose
the $k_s$ to transform to the `t Hooft scheme defined by $b'_\ell=0$ for $\ell
\ge 3$. For example, set $k_1=0$ and solve $b'_3=0$, obtaining $k_2
= b_3/b_1$; then solve $b'_4=0$, obtaining $k_3=b_4/(2b_1)$, and so forth for
higher $s$.
\bibitem{tanhgennote}
%
Eq. (\ref{tanh}) is the special case of Eq. (\ref{tanhgen}) with $r=4\pi$. For
$r \simeq 1$, we find that the ST (\ref{tanhgen}) can satisfy $C_1$-$C_4$ at 
an IRFP.
\bibitem{casimir}
%
The Casimir invariants $C_R$ and $T_R$ are defined as $\sum_a \sum_j {\cal
D}_R(T_a)_{ij} {\cal D}_R(T_a)_{jk} = C_R \delta_{ik}$ and $\sum_{i,j} {\cal
D}_R(T_a)_{ij} {\cal D}_R(T_b)_{ji} = T_R \delta_{ab}$, where $R$ is the
representation and $T_a$ are the generators of $G$, so that for SU($N_c$),
$C_A=N_c$ for the adjoint ($A$) and $T_{fund}=1/2$ for the
fundamental representation, etc. $C_f$ denotes $C_R$ for the fermion 
representation.
\bibitem{nfreal}
%
Here and elsewhere, when expressions are given for $N_f$ that evaluate
to non-integral real values, it is understood that they are formal and are 
interpreted via an analytic continuation of $N_f$ from physical integer
values.
\end{thebibliography}
\end{document}